\documentstyle[psfig, epsfig, 12pt]{article} 

\def\mypagenumber{1}
\def\myend{\end{document}}

\normalsize

\newcounter{sxn}

\newcounter{axn}

\date{}

\newdimen\mybaselineskip
\newcommand{\beq}{\begin{equation}}
\newcommand{\eeq}{\end{equation}}
\newcommand{\bea}{\begin{eqnarray}}
\newcommand{\eea}{\end{eqnarray}}
\newcommand{\ba}{\begin{eqnarray}}
\newcommand{\ea}{\end{eqnarray}}
\newcommand{\bpic}{\begin{picture}}
\newcommand{\epic}{\end{picture}}

\def\la{\raise.16ex\hbox{$\langle$} \, }
\def\ra{\, \raise.16ex\hbox{$\rangle$} }

\def\psibar{ \psi \kern-.65em\raise.6em\hbox{$-$} }
\def\mbar{ m \kern-.78em\raise.4em\hbox{$-$}\lower.4em\hbox{} }

\def\n@space{\nulldelimiterspace=0pt \mathsurround=0pt }
\def\huge#1{{\hbox{$\left#1\vbox to 20.5pt{}\right.\n@space$}}}

\def\myskip{\noalign{\kern 8pt}}
\def\myeqspace{\noalign{\kern 10pt}}

\def\boxit#1{$\vcenter{\hrule\hbox{\vrule\kern3pt
    \vbox{\kern3pt\hbox{#1}\kern3pt}\kern3pt\vrule}\hrule}$}
\def\bigbox#1{$\vcenter{\hrule\hbox{\vrule\kern5pt
     \vbox{\kern5pt\hbox{#1}\kern5pt}\kern5pt\vrule}\hrule}$}

\def\ignore#1{{}}

\begin{document}

\bibliographystyle{unsrt}
\footskip 1.0cm

\thispagestyle{empty}
\setcounter{page}{\mypagenumber}

             
\begin{flushright}{\today
\\ OUTP-00-19-P\\
 
}

\end{flushright}

\vspace{1.cm}
\begin{center}
{\Large \bf {Holographic renormalization group flow, 
}}\\ 
\vspace {.5cm}
{\Large  \bf {Wilson loops and field-theory $\beta$-functions}}\
\
\vspace{1cm}

{\bf 
Ian I. Kogan{\footnote{kogan@thphys.ox.ac.uk}}, 
Martin Schvellinger{\footnote{martin@thphys.ox.ac.uk}},
Bayram Tekin{\footnote{tekin@thphys.ox.ac.uk}}
}\\
\vspace{.5cm}
{\it Theoretical Physics, Department of Physics,
University of Oxford, 1 Keble Road, Oxford, OX1 3NP, UK}
\\
  
\end{center}

\vspace*{1.cm}


\baselineskip=15pt
We study the Renormalization Group (RG) flow of critical bosonic background fields
in the framework of the RG approach to string theory. 
In this approach quantum field theory $\beta$-functions are the extra inputs 
in solving the string theory sigma-model equations.
We study two different situations, the first one is the Yang-Mills theory 
where the coupling constant diverges in the infrared limit. 
The second case corresponds to a type of
theories where the $\beta$-function has a pole in the infrared limit and it changes
sign through the pole (as in ${\cal{N}}=1$ super-Yang-Mills theory). For this case
in the strong coupling branch, in the infrared, there is an interval of values of the coupling
in which the theory only leads to confinement. We have obtained this range. 
We also mention the theories with conformal-fixed points and their relation to theories
with a pole in the $\beta$-functions. We calculate the Wilson loops in these theories.  
\vfill



 
\newpage



\normalsize
\baselineskip=20pt plus 1pt minus 1pt
\parindent=25pt

\newpage
 
\section{Introduction}

The large $N$-limit of super-conformal gauge field theories has been proven 
to have dual representations in terms of the weakly coupled super-gravity or string theory 
via the Maldacena's conjecture \cite{MALDA,GUBSER,WITTEN,REPORT}. In such a duality the  
radius of the space $R_c$ behaves like $\Delta^{1/4}$, where
$\Delta=g_{YM}^2 N$ is the 't Hooft parameter which is fixed in the large-$N$ limit 
\cite{THOOFT}. Naturally for large $\Delta$ this leads to a large radius 
which allows us to describe strings propagating in the background fields \cite{CALLAN}.
This is a particular example of an idea suggested by Polyakov \cite{POL} about the 
description of gauge theories in $D$ dimensions in terms of certain non-critical string theory
in $D+1$ dimensions, where an extra dimension is due to the Liouville field. 
In this approach the properties of the geometry give us information about the properties of the 
gauge theory. On the other hand it should be possible to describe the weakly coupled regime of gauge theories 
by the strong coupling regime of string theory. 

One of the simplest questions that one can ask is about the derivation of the
RG flow in gauge theories from the geometry, including the weakly coupled regime. 
Here we are not going to deal with this in the full $10$-dimensional 
theory. Instead of that we are going to study an alternative approach due
to \'Alvarez and G\'omez \cite{AL} where the idea is to model the renormalization group
equations of gauge theories. Basically their proposal implies to associate to the couplings
of the gauge field theories some background fields of a closed string theory. Then one demands
that the string $\beta$-functions for those backgrounds have to coincide with the RG equations
of the gauge theories. In this approach the geometry dictates the properties of the gauge theory.
In particular, as it has been shown by \'Alvarez and G\'omez, for one-loop $\beta$-function
in pure gluodynamics the space-time curvature is a continuously increasing function of the
running-energy scale from the IR to the UV limit. 
The curvature behaves like an inverse power of the coupling.
It implies that the theory runs continuously from the strongly coupled regime to 
the weakly coupled one. In this framework they have calculated the 
Wilson loops and shown that both confinement and over-confinement occur depending only on the
area of the world-sheet of the fundamental string, at every value of the coupling.
In the present paper we address the following situation. 
Consider a theory with a $\beta$-function which has a pole at some energy scale $\Lambda$ in the infrared.
The existence of the pole leads to two
branches, one corresponds to the super-strongly coupled regime while the other one is
the asymptotically-free regime. The point $\Lambda$ behaves like an infrared attractive point
since the RG flow of the theory goes from the UV limit to the IR one in both branches \cite{KOGAN}.
Once the Wilson loops are computed for this kind of theories, one obtains an interesting new result:
in the super-strong coupling, in the infrared, there is a range of values of the coupling 
in which the theory only leads to confinement and not over-confinement. We have calculated that range. 

Starting from the bosonic string action it is possible to derive the one loop 
$\beta$-functions which yield the equations of motion of the background fields 
\bea
\beta^{\Phi} &=& \frac{D-26}{48 \pi^2} + \frac{{\alpha'}}{16 \pi^2} \left( 
  4 (\nabla \Phi)^2 - 4 \nabla^2 \Phi - R +\frac{1}{12} H^2 \right) +{\cal{O}}({\alpha'}^2) \,\,\, ,
  \nonumber \\
\beta^{G}_{\mu \nu} &=& R_{\mu \nu} - \frac{1}{4} H^{\lambda \sigma}_\mu H_{\nu \lambda \sigma} + 
  2 \nabla_\mu \nabla_\nu \Phi + {\cal{O}}({\alpha'}) \,\,\, , \nonumber \\
\beta^{B}_{\mu \nu} &=& \nabla_{\lambda} H^{\lambda}_{\mu \nu} - 2 (\nabla_\lambda \Phi) H^\lambda_{\mu \nu} + 
  {\cal{O}}({\alpha'}) \,\,\, , 
  \label{eq1} 
\eea
where $G_{\mu \nu}$ and $B_{\mu \nu}$ are the symmetric and antisymmetric fields, respectively.
$R$ is the scalar curvature and $R_{\mu \nu}$ is the Ricci tensor of the $D$-dimensional space-time.
$H_{\mu \nu \lambda}$ is the antisymmetric tensor-field strength derived from $B_{\nu \lambda}$.
These equations were derived using the world-sheet background-field perturbation theory
from the non-linear sigma-model action plus the renormalizable action for the dilaton field 
\beq
S_{dilaton}= \frac{1}{4 \pi} \int d \sigma \, \, d \tau \sqrt{\gamma} R^{(2)} \Phi(X) \,\,\, .
\eeq
Here $\gamma$ is the determinant of the 2-dimensional world-sheet metric, $R^{(2)}$ is the scalar
curvature of the world-sheet and $\Phi(X)$ is the dilaton background field in the $D$-dimensional target 
space-time $X$. In the low-energy theory instead of strings one can deal with the background fields.

Following \'Alvarez and G\'omez we will solve the equations of motion which come from the vanishing 
of the $\beta$-functions, Eq.(\ref{eq1}).
For simplicity we will turn off all the fields but the symmetric tensor and the dilaton. 
We will work in the critical dimension ($D=26$) for which the equations reduce to those derived from
the action of gravity coupled to the dilaton. 
Using the Liouville ansatz \cite{AL,POLYAKOV,POLYAKOV1} the equations of motion are trivially satisfied for
any dilaton. In the metric we identify
4 of the 26 dimensions as the usual Euclidean (or Minkowski) space-time where the gauge theory
is placed. One of the remaining $22$-spatial coordinates is identified as the running-energy
scale ($\mu$) associated to the gauge field theories. In this prescription the gauge coupling is 
related to the dilaton by using the association{\footnote {The second part of this relation is an assumption 
related to the soft-dilaton theorem \cite{AL}.}} $g_s = e^\Phi = g^2_{YM}$. 
Since we know how the coupling constants run in gauge theories through the $\beta$-functions, 
the above correspondence determines the $\mu$-dependence of the dilaton.
This procedure is known as the Renormalization Group approach to the string theory. See for example
\cite{AL,AKH,BAL,FRE,SAH} and references therein.

We describe some features of the space-time and calculate the Wilson loops.
There are two different types of $\beta$-functions that we will work on. The first is the one-loop
$\beta$-function of pure gluodynamics where the coupling constant diverges in the infrared. 
The second case is a $\beta$-function which changes sign through a pole in the infrared. 
For instance we will consider in particular the 
Novikov-Shifman-Vainstein-Zakharov (NSVZ) \cite{SHIFMAN}
$\beta$-function of ${\cal{N}}=1$ super-Yang-Mills theory, which has those properties.
It was conjectured that because of this pole this theory has two phases \cite{KOGAN}.
One is the super-strongly coupled phase and the other one is the asymptotically-free phase, both of 
which flow to the infrared attractive point at some finite scale. Finally in the discussions we mention 
about the theories with conformal-fixed points at the finite values of the coupling constants.

In section 2 we study the properties of the background fields and the corresponding confining
geometry derived from them. In section 3 the Wilson loops are studied by calculating the area of a fundamental 
string world-sheet in the background fields.
In sections 4 and 5 we apply this framework to investigate the Yang-Mills type 
$\beta$-function and the ${\cal{N}}=1$ super-Yang-Mills type one, respectively.

\section{The background fields}

Let us consider the RG equations of the bosonic strings \cite{CALLAN}. 
The vacuum configurations of string theory at one loop are determined 
by the sigma-model RG $\beta$-functions which,
at leading order in $\alpha'$, read as 
\beq
 \beta^G_{\mu \nu} = R_{\mu \nu} + 2 \nabla_\mu \nabla_\nu \Phi
                     \,\,\, ,
\eeq
and
\beq   
 8 \pi^2 \beta^\Phi = \frac{D-26}{6} - \alpha' \, \nabla^2 \Phi + 
            2 \alpha' (\nabla \Phi)^2  \, .
\eeq
Here $\nabla_\mu \Phi = \partial_\mu \Phi$ and 
$\nabla_\mu \nabla_\nu \Phi = \partial_\mu \partial_\nu \Phi - 
\Gamma^\alpha_{\mu \nu} \partial_\alpha \Phi$. 
Conformal invariance implies that $\beta^G_{\mu \nu} = \beta^\Phi = 0$. For all orders in 
$\alpha'$-ex\-pan\-sion these $\beta$-functions were shown to vanish \cite{PAF}. For the 
critical dimension these equations become 
\beq
 R_{\mu \nu} + 2 \, \nabla_\mu \nabla_\nu \Phi = 0  \,\,\, ,
\eeq
and
\beq
 \nabla^2 \Phi - 2 \, (\nabla \Phi)^2 = 0 \,\,\, .
\eeq 
As we have said before the four-dimensional space-time is embedded into $26$ dimensions and one of the remaining 
$22$-spatial coordinates is identified as the running-energy scale $\mu$. We will use the following form for 
the metric
\begin{equation}
 d s^2 = a(\mu) \,\, (\pm dt^2 + d {\vec{x}}^2) + b(\mu) \,\, d\mu^2 + c(\mu) \,\, d{\vec{y}}^2 \,\,\, ,
\end{equation}
where the $a(\mu)$ term is the Euclidean (Minkowski) metric in
four dimensions, $\mu$ is the fifth-coordinate, while
$d{\vec{y}}^2$ corresponds to an $21$-dimensional hyper-plane.
The Ricci tensor is computed to be
\beq
 R_{ij}= - \left( 
 \frac{a' b'}{4 b^2} - \frac{21 a' c'}{4 b c} -
        \frac{a'^2}{2 a b} - \frac{a''}{2 b} \right)\eta_{ij} \,\,\, ,
\eeq
for $i, \,\, j= 1, \,\, \cdot \cdot \cdot \,\, , \,\, 4$, where we choose the mostly
plus signature for the Minkowski metric, $\eta_{ij}=(-,+,+,+)$,
\bea
 R_{55} &=&  - \frac{a'^2}{a^2} - \frac{a' b'}{a b} + \frac{2 a''}{a} 
          - \frac{21 c'^2}{4 c^2} - \frac{21 b' c'}{4 b c} + \frac{21 c''}{2 c} \,\, ,  \nonumber \\
 R_{\alpha \beta}  &=& \left( \frac{a' c'}{a b} - \frac{b' c'}{4 b^2} + \frac{19 c'^2}{4 b c} 
                   + \frac{c''}{2 b} \right) \delta_{\alpha \beta}  \,\,\, ,
\eea
for $\alpha, \,\, \beta = 6, \,\, \cdot \cdot \cdot \,\, , \,\, 26$,
and the scalar curvature reads as
\beq
  R =   -\frac{2 b' a'}{a b^2} - \frac{21 b' c'}{2 b^2 c} + \frac{189 c'^2}{2 b c^2} 
   + \frac{a'^2}{a^2 b} + \frac{4 a''}{a b} + \frac{42 a' c'}{a b c} + \frac{21 c''}{b c} \,\,\, ,
\eeq
where prime denotes derivative with respect to $\mu$.
We will use the following ansatz to solve the equations of motion
\ba
a(\mu) & = & e^{2 \Phi} \,\,\, ,  \nonumber \\
b(\mu) & = & 4 e^{4 \Phi} \Phi'^2 \, , 
\label{ANSAT}
\ea
and $c(\mu)=1$. In this case the scalar curvature is $R=-e^{-4 \Phi}=-g_{YM}^{-8}$, where 
we have used $e^{\Phi}=g_{YM}^2$. 
There is a naked singularity in the space-time which corresponds to the weakly-coupled regime
of the gauge theory. On the other hand the
strongly-coupled regime of the gauge theory corresponds to the weakly-curved spaces. 
This allows us to make a straightforward analysis of the geometry in terms of the
quantum field theory $\beta$-functions. 

It is important to mention that we did not excite the tachyon field
here. In the general case the metric will be more complicated.
The geometry in this framework is universal in the sense that the metric and the scalar curvature depend only
on the coupling constant.  Choosing different types of theories, as long as the coupling diverges at certain point and goes to zero
at another one, corresponds to essentially choosing different coordinates in this geometry. However once the gauge
theory is introduced on a 4-dimensional
hyper-plane and the prescription of computing the Wilson loops in terms of the minimal surfaces 
is given, the flow of the coupling constant becomes important and therefore different theories, in principle, can behave differently.
With respect to this, and as we already mentioned before in the introduction, the important fact here is that the geometry 
can distinguish between one-loop $\beta$-function and $\beta$-functions with a pole at some finite value of the running-energy
scale.

\section{The Wilson loop}

In this section we discuss the calculation of the Wilson loops. First one should notice
that the metric which we are dealing with can be trivially rewritten just by using the ansatz given
in Eq.(\ref{ANSAT})
\begin{equation}
 ds^2 = e^{2 \Phi} (\pm dt^2 + dx_i dx_i) + 4 l_c^2 e^{4 \Phi} (d \Phi)^2  + 
 d{\vec{y}}^2 \,\,\, ,
 \label{METRIC}
\end{equation}
where $x_i$, $i=1, \, 2, \, 3$, and $l_c$ is an arbitrary scale of dimension of length.  
In this framework $l_c$ is related to the running-energy scale of the field theory studied\footnote{  
Moreover one can also define $\rho=e^{2 \Phi}$ and rewrite the above metric in the Liouville form
$ ds^2 = \rho \,\, (\pm dt^2 + dx_i dx_i) + l_c^2 \, (d \rho)^2  + 
         d{\vec{y}}^2$. In this parameterization $\rho=\infty$ is the horizon and $\rho$=0 is the singularity. Therefore
         this space-time allows the description of zigzag invariant Wilson loops \cite{POLYAKOV2}.}.
 
Since we want to deal with Wilson loops placed in the $4$-dimensional Euclidean (Minkowski) 
space-time, the extra 21 dimensions are irrelevant. Therefore
this problem is equivalent to the problem of $5$-dimensional 
gravity coupled to the dilaton.

Let us consider the Nambu-Goto action 
\beq
S_{NG}=\frac{1}{2 \pi l_s^2} \int d\sigma \, d\tau \sqrt{det G_{MN} \partial_\alpha X^M \partial_\beta X^N}
\,\,\, ,
\label{NAMBUGOTO}
\eeq
where $X^M$ is a generic coordinate on the $5$-dimensional space-time.
In the static configuration, for $\tau=t$ and 
$\sigma=x$ we have
\beq
 ds^2 = \pm \, e^{2 \Phi} \, d\tau^2 + 
 (e^{2 \Phi} + 4 \, l_c^2 \, e^{4 \Phi} \, \Phi_\sigma^2) \,\, d\sigma^2  \,\,\, ,
\eeq
where $\Phi_\sigma=\frac{\partial \Phi}{\partial \sigma}$. 
Therefore the action is
\beq
S_{NG}=\frac{T}{2 \pi l_s^2} \int^{L}_0 d\sigma \,  e^{2 \Phi} \sqrt{1 + 4 l^2_c e^{2 \Phi} \Phi_\sigma^2}
\,\,\, ,
\label{NEW}
\eeq
where $T$ is the time.
Basically we are considering the configurations as it is shown in figure 1.

\begin{center}
\epsfig{file=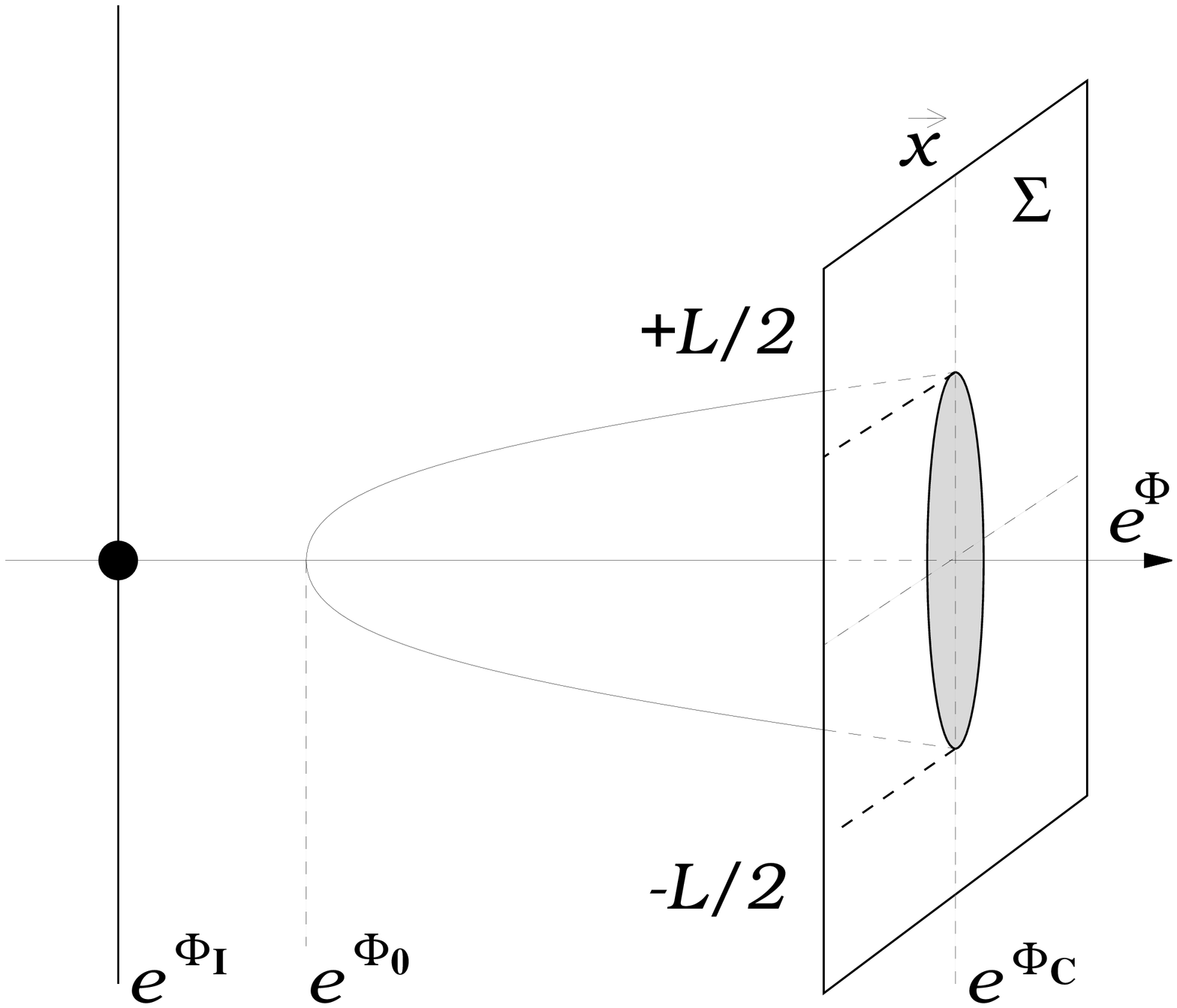, width=9.cm}
\end{center}
\baselineskip=13pt
\centerline{\small{Figure 1: Schematic representation of the prescription of calculating}}
\centerline{\small{the Wilson loops in terms of Liouville coordinates.}} 

\vspace{1.cm}

\baselineskip=20pt plus 1pt minus 1pt

In this figure we show the shape of the string world-sheet in terms of the
Liouville field $e^\Phi$. In particular $e^{\Phi_I}$ is the limit when $\Phi \rightarrow -\infty $. Since
the scalar curvature $R$ goes like $e^{-4 \Phi_I}$ it implies that at this point the metric has a naked singularity 
(the scalar curvature blows up). The point at the origin represents the naked singularity.
However in terms of the running-energy scale $\mu$ the location of this singularity 
depends on the particular field-theory $\beta$-function. We can put the $4$-dimensional
hyper-plane $\Sigma$ at any point which corresponds to choosing a particular value of the coupling constant.
Here we denote a generic choice by $e^{\Phi_c}$. Once the hyper-surface is chosen the string can fluctuate in either
of the directions orthogonal to $\Sigma$. However the minimal surfaces come from the strings that fluctuate in the
direction of the naked singularity as it is depicted in the figure. In terms of the gauge theory language this means that
semi-classically we allow the RG flow from strong coupling to the weak coupling regime. 
On the other hand, if an infrared attractive point occurs, the RG flow will be from the ultraviolet limit to the infrared one
\footnote{Observe that if one considers
the Liouville coordinate to be time-like with a metric of the form $ ds^2 = \rho \,\, (dt^2 + dx_i dx_i) 
- l_c^2 \, (d \rho)^2 $ 
the direction of RG flow is from weak coupling to strong coupling.}.
We will see that this is what happens for ${\cal{N}}=1$ super-Yang-Mills theory.

\vspace{0.5cm}
\begin{center}
\epsfig{file=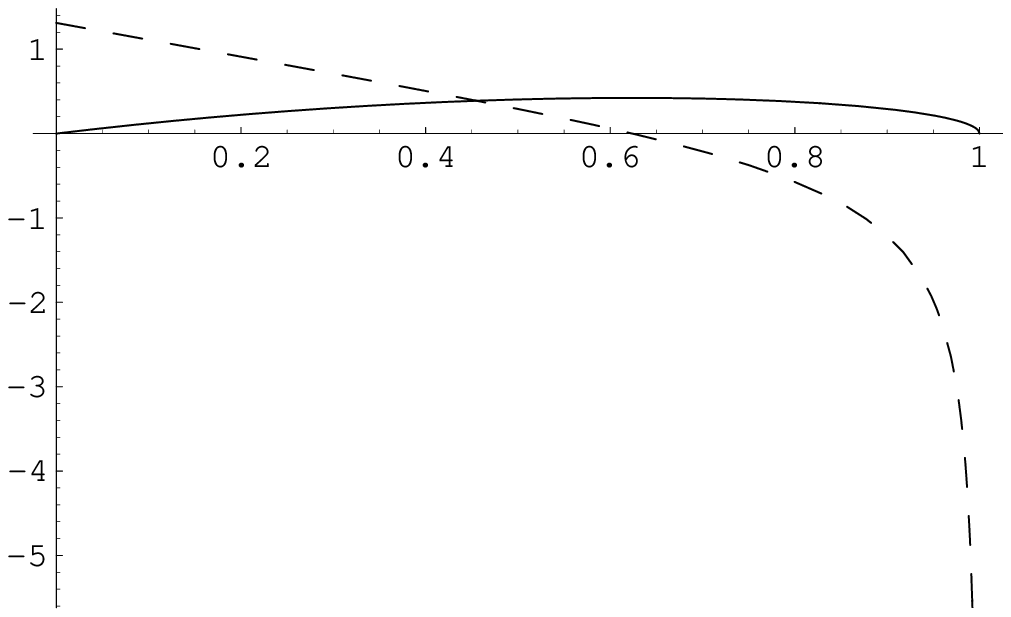, width=9cm}
\end{center}
\vspace{0.5cm}
\baselineskip=13pt
\centerline{\small{Figure 2: The solid line indicates the scaled Wilson-loop size $L/e^{\Phi_c}$ (in units of $l_c$)}} 
\centerline{\small{in terms of the dimensionless variable $e^{\Phi_0}/e^{\Phi_c}$. The dashed curve shows the}}
\centerline{\small{derivative of the scaled Wilson-loop size.}}
\vspace{0.5cm}

\baselineskip=20pt plus 1pt minus 1pt

Taking the vertical axis as the ${\vec{x}}$-direction we will consider a
Wilson-loop of size $L$. This means that $L$ is the static separation of the very heavy {"\it quark-anti-quark pair"}.
In the symmetric configuration ${\vec{x}}=0$ corresponds to a minimum  of $e^\Phi$, let us call it $e^{\Phi_0}$, and since the dilaton
is a function of $\mu$ we will call $\Phi_0=\Phi(\mu_0)$. For classical solutions one gets
\beq
\frac{e^{4 \Phi}}{1 + 4 l^2_c e^{2 \Phi} \Phi_\sigma^2}= e^{4 \Phi_0} \,\,\, .
\eeq
By inverting this expression 
\beq
\frac{L}{2} = 2 \, l_c \, e^{\Phi_0} \int^{e^{\Phi_c}/e^{\Phi_0}}_1 
\frac{1}{\sqrt{v^4-1}} \, dv \,\,\, ,
\eeq
where $v=\frac{e^{\Phi}}{e^{\Phi_0}}$. After integration we obtain
\beq
L = l_c \, e^{\Phi_0} \, \left( \sqrt{\pi} \frac{\Gamma(\frac{1}{4})}{\Gamma(\frac{3}{4})} - 4  
     \,\,\,\frac{_2F_1(\frac{1}{4}, \frac{1}{2}; \frac{5}{4}; e^{4 (\Phi_0 - \Phi_c})}{e^{\Phi_c-\Phi_0}}
\right) \,\,\, ,
\label{LLOOP}
\eeq
where $_2F_1(\frac{1}{4}, \frac{1}{2}; \frac{5}{4}; e^{4 (\Phi_0 - \Phi_c)})$ is a hypergeometric function \cite{ABRA}.
$L$ is a function of $e^{\Phi_c}$ and $e^{\Phi_0}$. 
It is convenient to study the function $L/e^{\Phi_c}$ in terms of $e^{\Phi_0}/e^{\Phi_c}$. It has
a maximum in the interval between $0$ and $1$ which indicates that there are two regions (I and II) to consider when one calculates the
Wilson loops \cite{AL}. For the interval below the maximum in the $e^{\Phi_0}/e^{\Phi_c}$-axis, region I,
one considers large world-sheets. In figure 2 we plot $L/e^{\Phi_c}$ in terms of $e^{\Phi_0}/e^{\Phi_c}$. There is
a maximum at $e^{\Phi_0^M}/e^{\Phi_c} \approx 0.62$ and at this point $L/e^{\Phi_c}|_{M}=0.42 \, \, l_c$.

Let us calculate the energy for the static configuration. The Nambu-Goto action 
and the corresponding energy are related by 
\beq
E=\frac{S_{NG}}{T}=\int {\cal{L}} \, d\sigma = \int {\cal{L}} \, \left( \frac{dv}{d\sigma} \right)^{-1} \, dv \,\,\, ,
\eeq
where the lagrangian is given by
\beq
{\cal{L}}= \frac{e^{2\Phi}}{2 \pi l^2_s} \sqrt{1 + 4 l_c^2 e^{2 \Phi} \Phi^2_\sigma} = 
           \frac{1}{2 \pi l^2_s} e^{2\Phi_0} v^4 \,\,\, .
           \label{LOOP1}
\eeq
Then the integral becomes
\beq
E = \frac{2 l_c}{\pi l^2_s} \, e^{3 \Phi_0} \, \int_1^{\frac{e^{\Phi_c}}{e^{\Phi_0}}} \, 
    \frac{v^4}{\sqrt{v^4 - 1}} \, dv \,\,\, ,
\label{EE}
\eeq
which can be integrated as 
\beq
E= \frac{ 2 l_c}{\pi l^2_s} \,\,\, e^{3 \Phi_0} \,\,\, \left( \frac{\sqrt{\pi}}{12} \,
   \frac{\Gamma(\frac{1}{4})}{\Gamma(\frac{3}{4})} + \frac{1}{3} e^{3(\Phi_c-\Phi_0)} 
   \,\,\, _2F_1 \left( \frac{1}{2}, \frac{-3}{4}; \frac{1}{4}; e^{4 (\Phi_0 - \Phi_c)} \right) \right) \,\,\, .
   \label{ENERGY1}
\eeq
Expanding Eqs.(\ref{LLOOP}) and (\ref{ENERGY1}) in series of 
powers of $e^{\Phi_0}/e^{\Phi_c}$ we obtain
\beq
\frac{L}{l_c}  =  \sqrt{\pi} \, e^{\Phi_0} \, \frac{\Gamma(\frac{1}{4})}{\Gamma(\frac{3}{4})} - 4 \, 
                    e^{2\Phi_0-\Phi_c} + {\cal{O}}(e^{6\Phi_0-5\Phi_c})
  \,\,\, ,
  \label{LL2}
\eeq
and
\beq
\frac{\pi l^2_s E}{2 l_c}  =  \frac{e^{3 \Phi_c}}{3} + \frac{\sqrt{\pi}}{12} \, \frac{\Gamma(\frac{1}{4})}{\Gamma(\frac{3}{4})} 
                    \, e^{3 \Phi_0}- \frac{1}{2} \, e^{4\Phi_0-\Phi_c} + {\cal{O}}(e^{8\Phi_0-5\Phi_c})
  \,\,\, .
\label{ENERGY2}
\eeq
Replacing $L$ in terms of $e^{\Phi_0}$ in Eq.(\ref{ENERGY2}) we get an expression which shows that there is
over-confining.
\beq
E = \frac{1}{6 \pi^2 l^2_s l^2_c} \, \left( \frac{\Gamma(\frac{3}{4})}{\Gamma(\frac{1}{4})} \right)^2 \, \, L^3 +
   \frac{2}{3 \pi} \frac{l_c}{l^2_s}   e^{3 \Phi_c} \,\,\, .
    \label{gap}
\eeq
In the limit of extremely strong coupling, which corresponds to $e^{\Phi_c} \rightarrow \infty$, the above expression becomes
divergent. However the divergent part is independent of the size of the Wilson loop. 
In the context of AdS/CFT duality this kind of divergence was regularized by a mass renormalization \cite{MALDAWILSON} or by considering
the Legendre transform of the minimal area \cite{GROSS}.   
In our case observing that zero-size Wilson loops diverge we will drop the $L$-independent divergent term  
and measure the energy with respect to the zero-size Wilson loops. 
In the extremely strong coupling limit $L$ becomes
\beq
L_{\infty} = \sqrt{\pi} \, l_c \, e^{\Phi_0} \, \frac{\Gamma(\frac{1}{4})}{\Gamma(\frac{3}{4})} \,\,\, ,
\eeq
while the energy for the static configuration reads as follows
\beq
E_{\infty} = 
 \frac{l_c \, e^{3 \Phi_0}}{6 \sqrt{\pi} l^2_s} \frac{\Gamma(\frac{1}{4})}{\Gamma(\frac{3}{4})} \,\,\, ,
\eeq
so that the over-confining is shown by the relation
\beq
E_{\infty} = \frac{1}{6 \pi^2 l^2_s l^2_c} 
             \frac{\Gamma(\frac{3}{4})^2}{\Gamma(\frac{1}{4})^2} L^3_{\infty} \,\,\, .
\eeq
\begin{center}
\epsfig{file=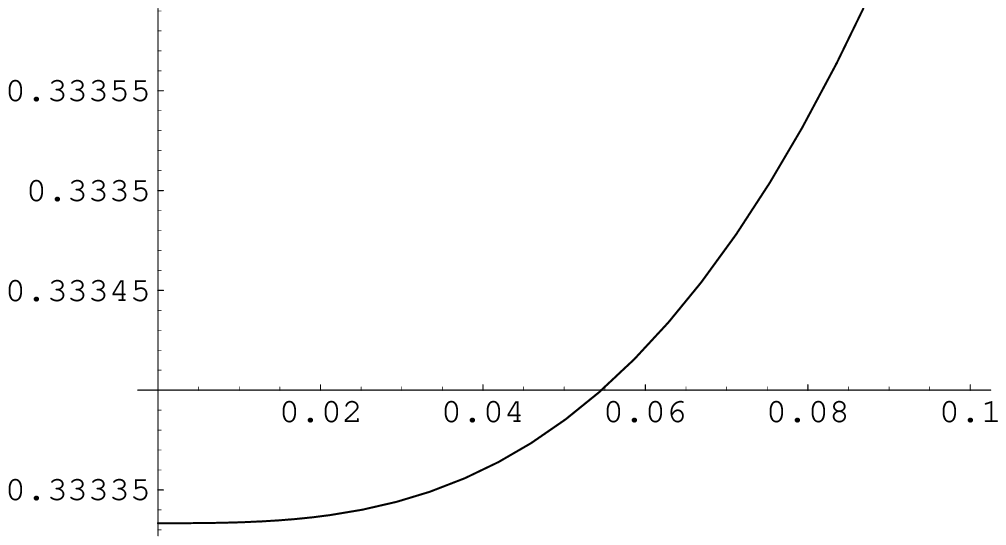, width=9cm}
\end{center}
\baselineskip=13pt
\centerline{\small{Figure 3: Energy (in units of $\frac{l_c}{l_s^2}$) for the static configuration}} 
\centerline{\small{ as a function of $\frac{e^{\Phi_0}}{e^{\Phi_c}}$, in a small interval around $0$.}}
\vspace{1.cm}
\baselineskip=20pt plus 1pt minus 1pt
In figure 3 we plot the energy versus $e^{\Phi_0}/e^{\Phi_c}$
in a small interval around zero (region I). It shows the cubic behaviour of the potential.
A quite different situation arises when the region II is analyzed.  In this region
$e^{\Phi_0}$ is close to $e^{\Phi_c}$. In such a case we have to do the integration between
$1$ and $1+\epsilon$ so that we have
\beq
\frac{L_\epsilon}{2}  = 
  2 l_c \, e^{\Phi_0} \,  \left( \sqrt{\pi} \, \frac{\Gamma(\frac{5}{4})}{\Gamma(\frac{3}{4})} -
  (1+\epsilon) \, \, _2F_1 \left(\frac{1}{4}, \frac{1}{2}; \frac{5}{4}; 
  \frac{1}{(1+\epsilon)^4}\right) \right) = 
  2 l_c \, e^{\Phi_0} \, \left( \sqrt{\epsilon} + {\cal{O}}(\epsilon^{\frac{3}{2}}) \right) \,\,\, ,  
\eeq
and
\bea
E_\epsilon  &=&  
  \frac{2 l_c}{\pi l^2_s} \, e^{3 \Phi_0} \,  \left( \frac{\sqrt{\pi}}{12} \, \frac{\Gamma(\frac{1}{4})}{\Gamma(\frac{3}{4})} + 
  \frac{(1+\epsilon)^3}{3} \, _2F_1 \left(\frac{1}{2}, \frac{-3}{4}; \frac{1}{4}; 
  \frac{1}{(1+\epsilon)^4}\right) \right) \nonumber \\
&=&   \frac{2 l_c}{\pi l^2_s} \, e^{3 \Phi_0} \, \left( \sqrt{\epsilon} + {\cal{O}}(\epsilon^{\frac{3}{2}}) \right) \,\,\,  .
\eea
At order $\sqrt{\epsilon}$ it gives 
\beq
E_\epsilon = \frac{e^{2 \Phi_0}}{2 \pi l^2_s} \, L_\epsilon \,\,\, ,
\label{LINEAL}
\eeq
indicating confinement at region II. 
Figure 4 shows the whole picture of $\frac{E}{e^{3 \Phi_c}}$ (solid line) and $\frac{L}{e^{\Phi_c}}$ (dashed line), as functions of
$e^{\Phi_0}/e^{\Phi_c}$. In the region II we observe the area law ($S \approx L T$). 
\vspace{0.7cm}
\begin{center}
\epsfig{file=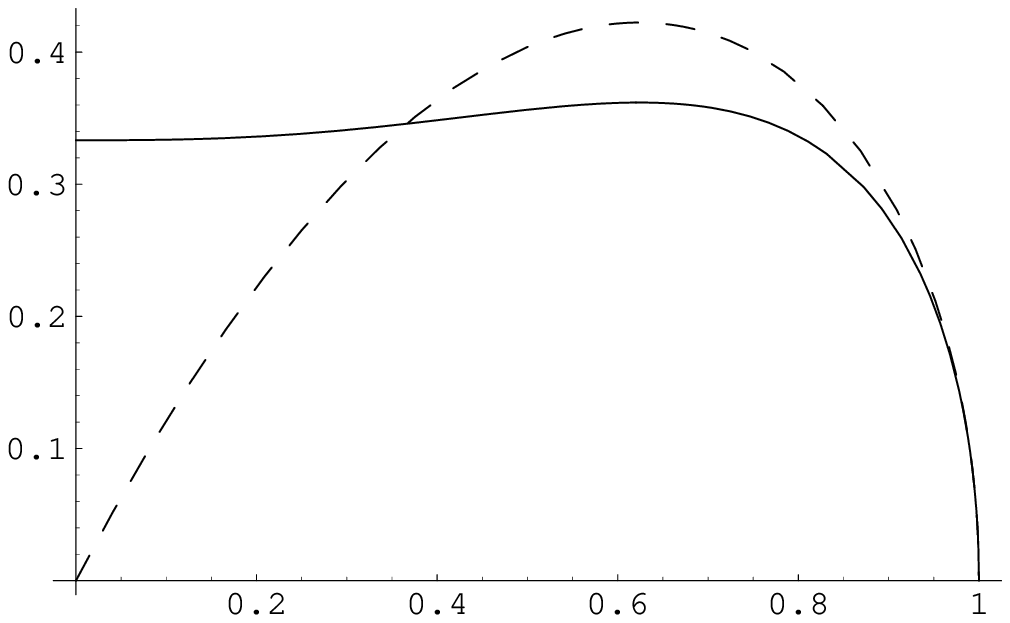, width=9cm}
\end{center}
pt
\centerline{\small{Figure 4: $\frac{E}{e^{3 \Phi_c}}$ (solid line) and $\frac{L}{e^{\Phi_c}}$ (dashed line),}}
\centerline{\small{as functions of $e^{\Phi_0}/e^{\Phi_c}$.}}
\vspace{0.7cm}
\baselineskip=20pt plus 1pt minus 1pt

\section{One-loop $\beta$-function}

In this section we will study the one-loop $\beta$-function of pure gluodynamics and compute the Wilson loops
in this theory. This case was studied in \cite{AL}. The $\beta$-function is
 \beq
 \mu \frac{d g}{d \mu} = - \frac{11 N}{24 \pi^2} g^3 \,\,\, .
\eeq
The dilaton field is
\beq
\Phi (\mu) = - \log \log \left( \frac{\mu}{\Lambda} \right) 
             - \log \left( \frac{11 N}{24 \pi^2} \right) \,\,\, ,
\eeq
where $\Lambda$ is a renormalization scale. The solutions for the metric are 
\beq
a(\mu) = \frac{576 \pi^4}{121 N^2} \frac{1}{\log^2 (\mu/\Lambda)} \,\,\, ,
\eeq
and
\beq 
b(\mu) = \frac{3^4 \, 2^{14} \pi^8}{(11 N)^4} \frac{1}{\mu^2 \log^6(\frac{\mu}{\Lambda})} \,\,\, .
\eeq
The curvature reads as
\beq
 R(\mu) = - \left( \frac{11}{3} \right)^4 \frac{N^4}{2^{12} \pi^8} \log^4 (\mu/\Lambda) \,\,\, .
\eeq
Figure 5 shows a picture for the Yang-Mills coupling constant in terms of the running energy scale $\mu$. 
\vspace{0.7cm}
\begin{center}
\epsfig{file=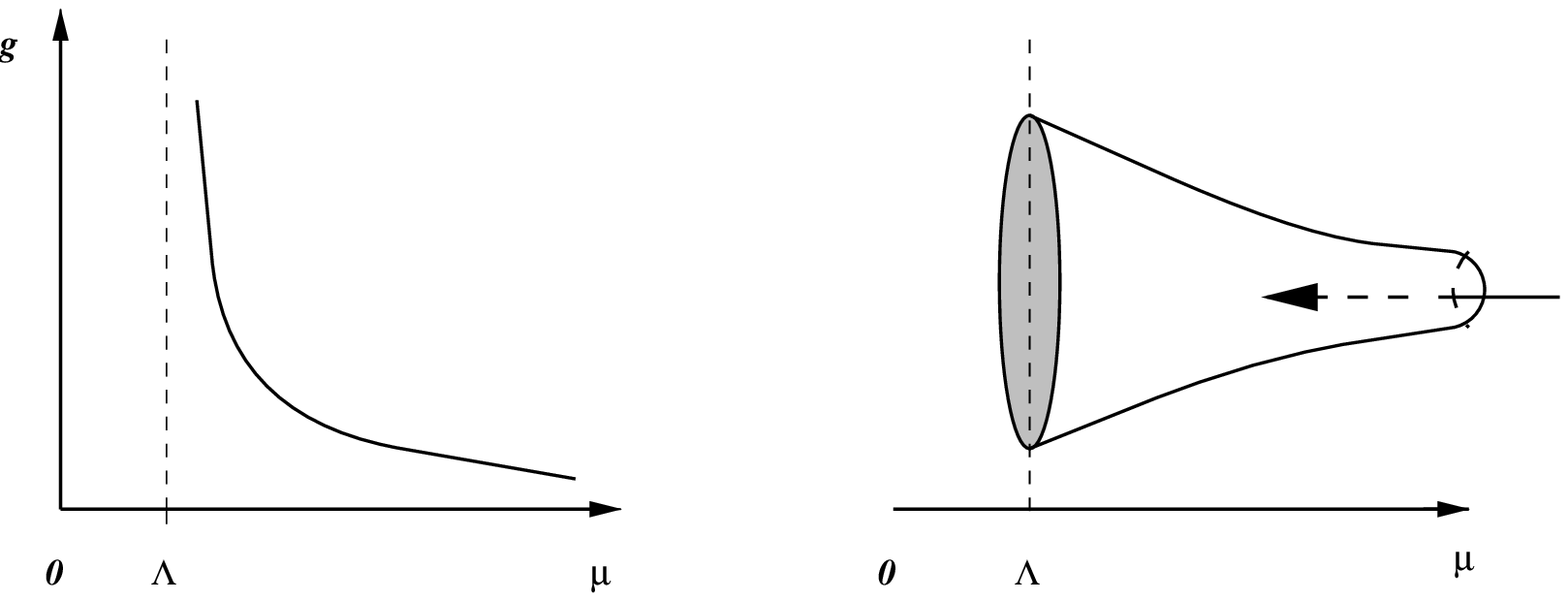, width=13cm}
\end{center}
\baselineskip=13pt
\centerline{\small{Figure 5: Coupling constant for pure gluodynamics as a function of the}}
\centerline{\small{running-energy scale $\mu$. The picture on the right shows the}}
\centerline{\small{qualitative behaviour of the four-dimensional space-time.}}
\vspace{1.cm}
\baselineskip=20pt plus 1pt minus 1pt
The picture on the right shows the behavior of the space-time radius (inverse of the curvature) as
$\mu$ approaches to the IR limit. At $\mu=\Lambda$ the space-time becomes flat. 
In the weak coupling there is a naked singularity. 

We will use the previous analysis for the Wilson loop calculation.  
In figure 6 we show the shape of the Wilson loop. The $4$-dimensional hyper-plane
$\Sigma$ where the Wilson loop is drawn can be placed at any value of $\mu$, let us say $\mu_c$, while
the minimum $e^{\Phi_0}$ corresponds to $\mu_0$. The naked singularity, represented by a dot, is the point at infinity
and it is labeled as $\mu_I$. The corresponding point to the maximum in figure 4, {\it i.e.}
$\frac{e^{\Phi^M_0}}{e^{\Phi_c}}$, is for this case 
$\mu^M_0 \approx \Lambda \left(\frac{\mu_c}{\Lambda}\right)^{1.61}$. For larger world-sheets it
leads to over-confinement 
\beq
E = \frac{1}{6 \pi^2 l^2_s l^2_c} \, \left( \frac{\Gamma(\frac{3}{4})}{\Gamma(\frac{1}{4})} \right)^2 \, \, L^3 
     \,\,\, . \nonumber
    \label{gapqcd}
\eeq
Note that we have dropped the divergent part. While for smaller world-sheets we get
\beq
E_\epsilon = \frac{288 \pi^3}{121 N^2 \, l_s^2 \, \log^2(\frac{\mu_0}{\Lambda})} \, L_\epsilon \,\,\, ,
\label{LINEALQCD}
\eeq
indicating the confinement in region II. The factor in front of $L_\epsilon$ can be interpreted as an effective tension.

\begin{center}
\epsfig{file=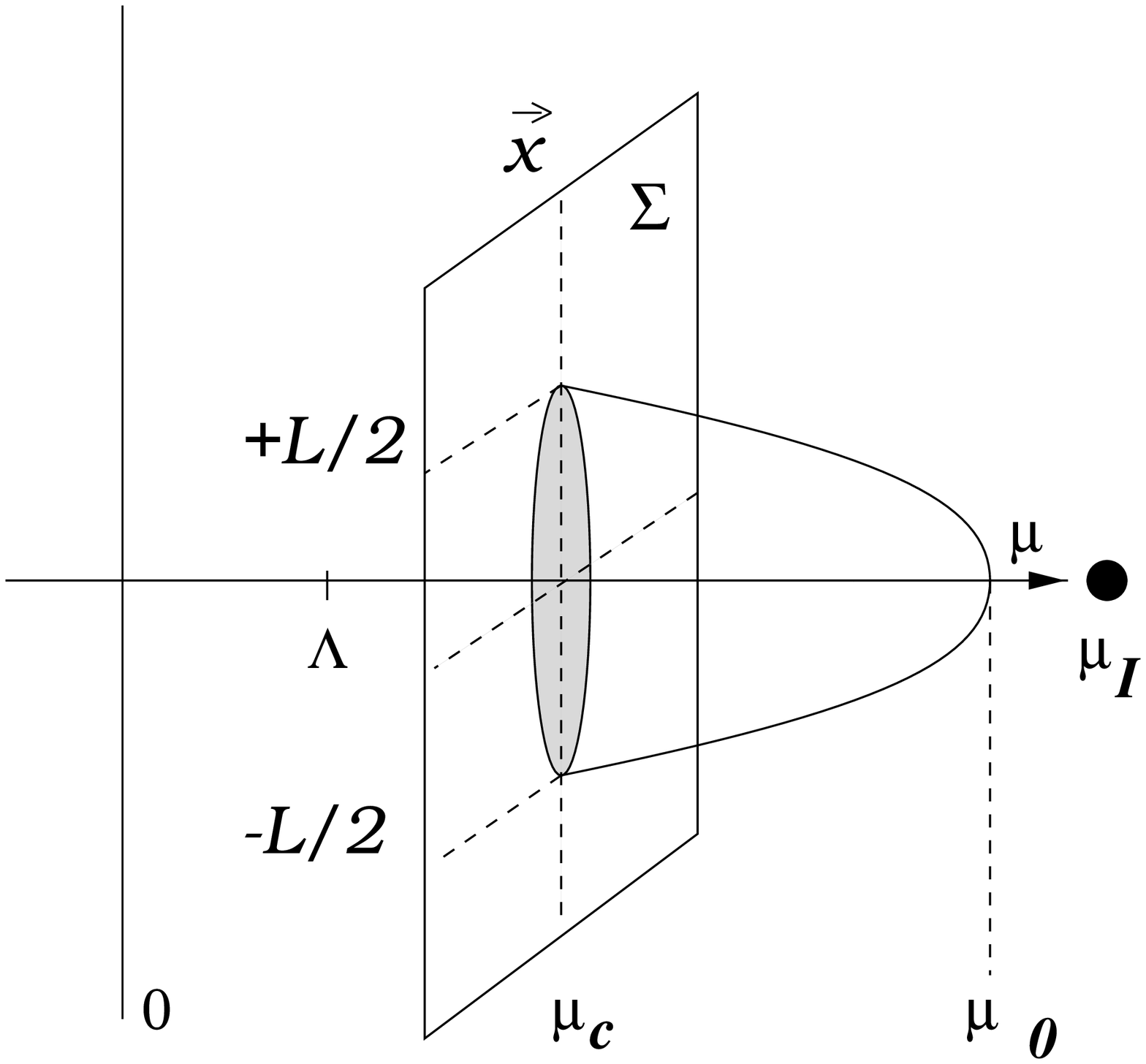, width=9cm}
\end{center}
\centerline{\small{Figure 6: Schematic representation of the Wilson loop calculation}}
\centerline{\small{for pure gluodynamics.}}
\vspace{0.7cm}

\section{$\beta$-function with a pole}

On the other hand, for a $\beta$-function with a pole at some finite value of the coupling
the situation is quite different. For instance let us
consider the ${\cal{N}}=1$ super-Yang-Mills theory. In this case we will study  
NSVZ $\beta$-function \cite{SHIFMAN}
\beq
\beta(g) = \mu \frac{d g}{d \mu} = 
 \frac{g^3}{16 \pi^2} \frac{({\cal{N}}-4)N}
 {\left( 1-\frac{(2-{\cal{N}})N g^2}{8 \pi^2}\right)} \,\,\, ,
\eeq
where $N$ is the corresponding label to the gauge group $SU(N)$ while ${\cal{N}}$ 
labels the number of super-symmetries. 
By integrating this equation one gets a transcendental equation
\beq
 {\mu\over\Lambda} = e^{8 \pi^2 \over{(4-{\cal{N}}) \, \, N g^2}} \,\,
                          \left({g^2\over {4 \pi}} \right)^{{2-{\cal{N}}}\over{4-{\cal{N}}}} \,\,\, .
\label{trans}                       
\eeq
For ${\cal{N}}=1$ super-Yang-Mills theory the $\beta$-function is exact both perturbatively and non-perturbatively and it reads 
\begin{equation}
 \beta(g) = \mu \frac{d g}{d \mu} = 
 - \frac{3}{16 \pi^2} \frac{g^3 N}{\left( 1-\frac{N g^2}{8 \pi^2} \right)} \,\,\, .
\end{equation}
This $\beta$-function has a pole at  $g^2 = 8\pi^2/N$ and it changes sign through the pole. As it is shown in figure 7
the coupling constant is a double-valued function of $\mu$. The pole is an infrared attractive point. 

\begin{center}
\epsfig{file=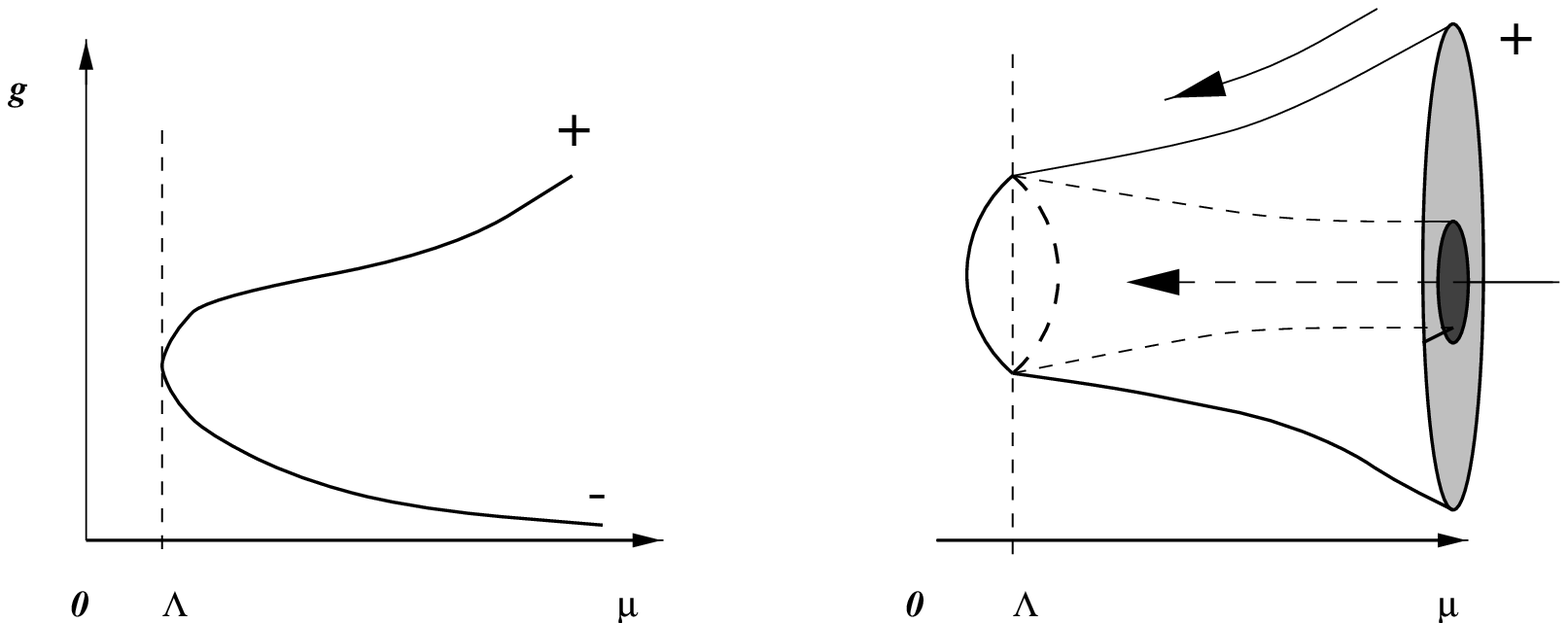, width=13cm}
\end{center}
\baselineskip=13pt
\centerline{\small{Figure 7: Coupling constant for ${\cal{N}}=1$ super-Yang-Mills theory as a}}
\centerline{\small{function of the running-energy scale $\mu$. The picture on the right shows}}
\centerline{\small{the qualitative behaviour of the four-dimensional space-time.}}
\vspace{1.cm}

\baselineskip=20pt plus 1pt minus 1pt

The theory can flow, both from the asymptotically-free phase where the coupling is small at large $\mu$ (lower branch ($-$)), 
and from the super-strongly coupled phase where the coupling is large (upper branch ($+$)), 
to the infrared attractive point. Since we can not invert the transcendental equation (\ref{trans}) we will 
analyze the behaviour around the infrared point $\Lambda$. By expanding the Eq.(\ref{trans}) around $\Lambda$ one gets
\beq
 g^2_{\pm} = 
 \frac{8 \pi^2}{N} \left( 1 \pm \sqrt{3 \frac{\mu-\Lambda}{\Lambda}} \right) \,\,\, .
\eeq
Since $\Phi = \log (g^2)$ it follows that
\beq
 \Phi_{\pm} = \log (8 \pi^2/N) + \log(1 \pm \zeta) \,\,\, ,
\eeq
where $\zeta = \sqrt{3 (\mu-\Lambda)/\Lambda}$. The solutions are
\ba
 a_{\pm}(\mu) &=& \frac{64 \pi^4}{N^2} (1 \pm \zeta)^2            \,\,\, ,
\nonumber \\
 b_{\pm}(\mu)&=& 9 \frac{ 2^{12} \pi^8 }{N^4 \Lambda^2}   \frac{(1 \pm \zeta)^2}{\zeta^2} \,\,\, .
\ea  
We set $c_\pm(\mu)= 1$. The scalar curvature is given by
\ba
 R_{\pm}&=& -\frac{N^4}{ 2^{12} \pi^8 ( 1 \pm \zeta)^4} \,\,\,\,\,\,\,\,\,\,\,\,\,\,\,\,\,\, \zeta << 1 \,\,\, .
\ea

In figure 8 we show the picture of the Wilson loop for the present case.
Notice that we use the same coordinates as in figure 1. In these coordinates the picture becomes
more clear. 

The vertical line at the point $e^{\Phi_\Lambda}$ represents the infrared attractive point.
Here it acts like an effective horizon. This is because in the super-strongly coupled phase
the direction of the RG flow is from strong coupling to the 
weak coupling, however since we have an infrared attractive point we can not continue the flow to the smaller couplings.
On the other hand in the asymptotically-free phase the flow is from weak coupling to the strong coupling.
The singularity placed at the origin corresponds to the limit of weak coupling shown in the lower-branch
of figure 7. Again confinement and over-confinement have to be understood as properties of the world-sheet size. 
Close to effective horizon, for both branches we have
\beq
E_{\epsilon \pm}= \frac{32 \pi^3 \left( 1 \pm \sqrt{3 (\mu_0-\Lambda)/\Lambda} \right)^2}{N^2 l_s^2} \, L_\epsilon \,\,\, .
\label{LINEALq}
\eeq

In figure 8 the left side of the vertical line represents the weakly-coupled phase 
and the right side represents the super-strongly coupled phase.

\vspace{0.5cm}
\begin{center}
\epsfig{file=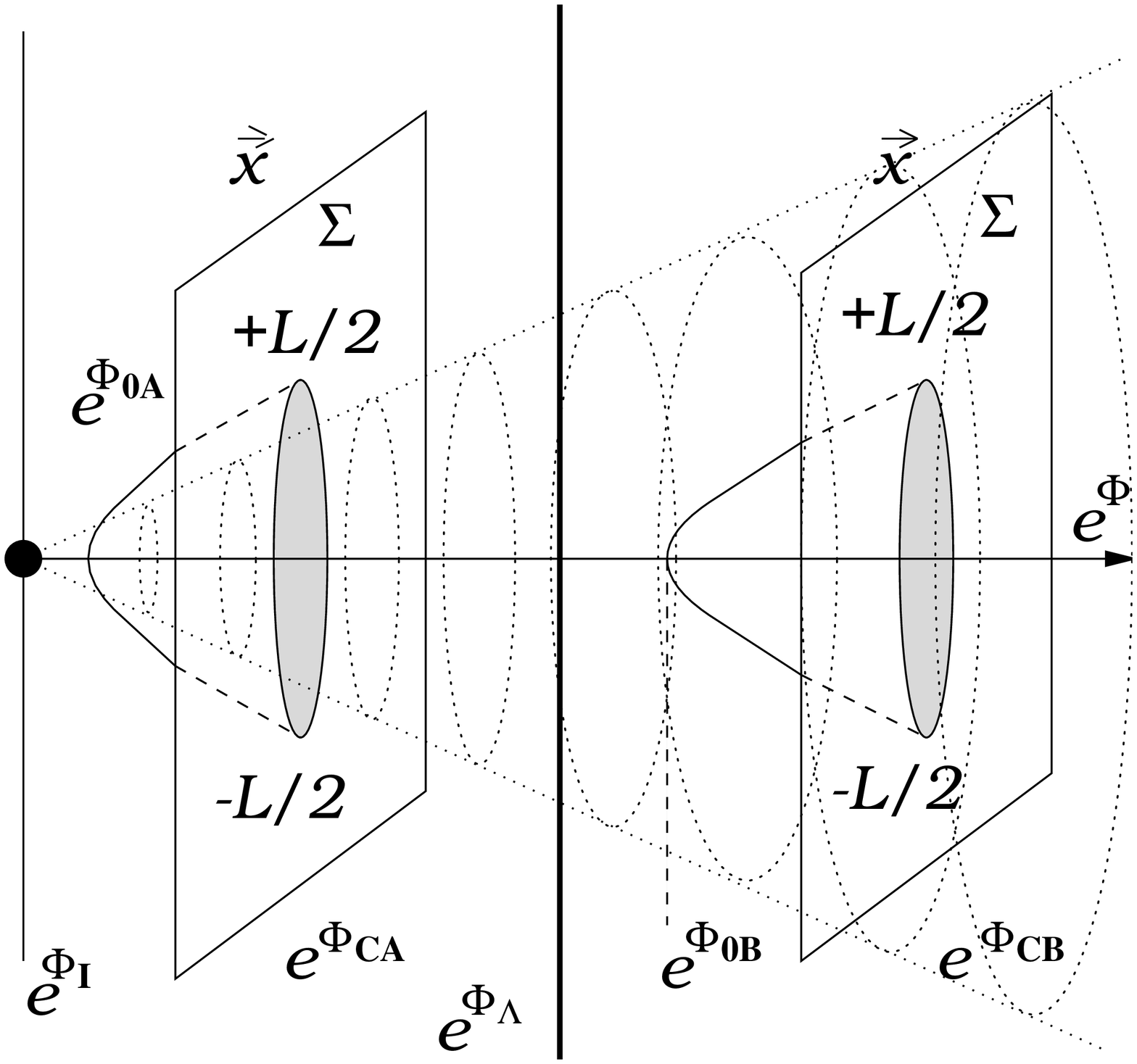, width=9cm}
\end{center}
\baselineskip=13pt
\centerline{\small{Figure 8: Schematic representation of the Wilson loop for ${\cal{N}}=1$ super}}
\centerline{\small{Yang-Mills theory. The dashed ellipses represent the behaviour of the}}
\centerline{\small{$4$-dimensional space-time as it has seen in the right picture of figure 8.}}
\vspace{0.5cm}

\baselineskip=20pt plus 1pt minus 1pt

When one studies the theory in the super-strongly coupled phase and sufficiently close to the 
attractive point one always gets the area law for the Wilson loops. In fact there is a critical
coupling $g_{critical}= 1.27 \, g_\Lambda$. For all values of the coupling between $g_{critical}$ and $g_\Lambda$
we have the area law.

\section{Discussion and conclusions}

In this paper, using the RG approach to string theory, we studied two types of confining theories.
As it was shown by \'Alvarez and G\'omez \cite{AL} a theory which has a $\beta$-function with a 
zero at UV-limit is confining or over-confining, depending on the minimal surfaces we use. 
We have considered the theories which have $\beta$-functions with
a pole at some finite value of the coupling. For example ${\cal{N}}= 1$ super-Yang-Mills theory is of this form. It has
two phases, the super-strongly coupled phase and the asymptotically-free phase which flow to an infrared attractive point.
In this theory one has also both confinement and over-confinement. We showed that in the super-strongly 
coupled regime of the theory there is a critical value below which one has confinement only. In this regime
large world-sheets are not allowed and one does not have over-confinement. It is interesting to stress that
this fact is due to the existence of a pole at some finite scale in the $\beta$-function. This leads to a new scale
$g_{critical}=1.27 \, g_\Lambda$ below which these theories satisfy the area law for their Wilson loops.

Concerning to the flow of the space-time for the one-loop $\beta$-function we have seen that it goes from the
strong coupling to the weak coupling regime. In the case of $\beta$-functions with a pole the situation changes
in the sense that now the RG flow goes from the UV limit to the IR limit, in both branches. 

The geometry we have considered
is universal in the sense that it leads to confinement. So presumably this geometry is not suitable 
for conformal theories and Abelian theories. For the conformal case we know that the space-time should be
anti-de Sitter. However let us assume that, by using the metric in this paper, we can describe theories
with conformal-fixed points. The $\beta$-function changes sign through zero instead of a pole but 
the geometry and the analysis of the Wilson loops will be similar to the case of $\beta$-functions with a pole which was 
discussed in section 5. Although this is surprising, an argument given by
Damgaard and Haangensen \cite{DAMGAARD} may help one to understand this. They showed a theory
with a self-dual point is either conformal at the self-dual point or its $\beta$-function
has a pole at this point. Here we reproduce their argument. Let us consider a theory with a coupling $g$.
If this theory is self-dual there is a relation between $g$ and the coupling of its dual theory $g^*$ 
\beq
g^*=f(g) \,\,\, .
\label{MAP}
\eeq
The only essential requirement is that the interaction part in the action and its dual have to be of
the same form. The map in Eq.(\ref{MAP}) is assumed to be {\it one-to-one} and
$f \cdot f =1$, which implies that its derivative $\frac{\partial f}{\partial g}$ is negative.
We also assumed that $g$ and $g^*$ are positive-valued functions. At the self-dual point, let us call it
$g_{self}$ we have 
\beq
g_{self}=f(g_{self}) \,\,\, .
\eeq
On the other hand we know that the RG flow is dictated by the $\beta$-function
\beq
\beta(g) = \mu \frac{\partial g}{\partial \mu} \,\,\, , \nonumber
\eeq  
by applying the operator $\mu \frac{\partial }{\partial \mu}$ to the map of Eq.(\ref{MAP})
we obtain a consistency relation
\beq
\mu \frac{\partial g^*}{\partial \mu} = \mu \frac{\partial g}{\partial \mu} \frac{\partial g^*}{\partial g} = 
  \beta(g) \, \frac{\partial f}{\partial g} \,\,\, ,
  \label{CONSIST}
\eeq
if one identifies the left hand side of the above equation with the $\beta$-function for the dual theory
one obtains
\beq
\beta(g^*) = \beta(g) \, \frac{\partial f}{\partial g} \,\,\, .
\eeq 
From the condition $f \cdot f=1$ at the self-dual point results that $\frac{\partial f}{\partial g}=-1$,
and therefore there are two solutions of Eq.(\ref{CONSIST}), {\it i.e.} the $\beta$-function can be zero at
this point or it is discontinuous. 
The most natural way to realize the discontinuity is through a pole in the $\beta$-function.
Then one has at the self-dual point $\beta(g_{self}+\epsilon) = -\beta(g_{self}-\epsilon)$.

Finally we would like to speculate that when one  deforms ${\cal{N}}= 4$ theory down to ${\cal{N}}= 1$ to
obtain a dual description in terms of string theory or gravity, a prescription  which works only in the strong coupling 
regime of the gauge theory, 
one might be dealing  with the super-strongly coupled regime, the upper branch, of ${\cal{N}}= 1$ theory.  

\section*{Acknowledgments}

We benefited from useful discussions with Alex Kovner. M.S. also acknowledges 
discussions with Gast\'on Giribet, Esteban Moro and Carlos N\'u\~{n}ez, 
and the organizers of the Spring Workshop on Superstrings and 
Related Matters (2000) at the Abdus Salam ICTP for kind hospitality where a part 
of this work was carried out. The work of I.K. and B.T. 
was supported by PPARC Grant PPA/G/O/1998/00567. 
The work of M.S. was supported by the 
CONICET of Argentina, the Fundaci\'on 
Antorchas of Argentina and The British Council.

\vfill

\end{document}